\documentclass[showpacs,aps,prl]{revtex4}
\usepackage{graphicx}%
\usepackage{dcolumn}
\usepackage{amsmath}
\usepackage{latexsym}
\begin{document}
\title{Can there be a Fock state of the radiation field?}
\author{N. Nayak\footnote{E-Mail: nayak@bose.res.in}}
\affiliation {S. N. Bose National Centre for Basic Sciences, 
Block-JD, Sector-3, Salt Lake City, Kolkata-700098, INDIA}
\begin{abstract}
We analyze possible hurdles in generating a Fock state of the radiation field in 
a micromaser cavity.
\end{abstract}
\maketitle
We plan to answer this question with the Munich micromaser in mind [1]. It 
consists of a superconducting cavity maintained at $T=0.3 K$. 
Hence, the average thermal photons present in the cavity is $\bar n_{th}=0.033$. 
The cavity dissipation parameter $\kappa=\nu/2Q$ stands at 3.146 Hz with the cavity 
$Q=3.4\times 10^9$ and $\nu$ being the masing frequency.
A clever velocity selector sends $^{85}Rb$ atoms in the upper of its two 
Rydberg levels into the cavity at such a rate that at most one atom is present there at 
a time. Also, the velocity selector maintains a constant flight time for each and 
every atom through the cavity. This is crucial for the Jaynes-Cummings [2] 
interaction between the single mode of the cavity and the atom present there. The 
attempt is to generate a Fock state of the cavity radiation field. To start with, 
the cavity is in thermal equilibrium having the normalized variance
\begin{equation}
 v=\sqrt{(\langle n^2\rangle -\langle n\rangle ^2)/\langle n\rangle}=\sqrt{1+{\bar n_{th}}}.
\end{equation}
The cavity at $T=0.3 K$ has $v=1.0164$. The evolution of $v$ has to be from this value to 
zero if one plans to generate a photon Fock state in the cavity. \\ \\
\indent
Our earlier anlysis [3] indicated such a possibility if and only if $\bar n_{th}=0$, 
that is, the cavity temperature has to be at $T=0 K$, a feat unattainable experimentally. 
However, the theory there followed an iterative procedure. 
Surely, we have to adopt an exact procedure in order to get a correct answer to the 
question in the title of this letter. Further, the reservoir effects have to be 
properly addressed to since the Fock states are very amneble to the dissipative forces. 
For this reason we find other approaches in the literature [4] unsuitable for the 
present purpose since the cavity dissipation is completely neglected ($Q=\infty$) 
there during the atom-field interaction. Hence, we look for a solution of the equation 
of motion\\ 
\begin{eqnarray}
\dot{\rho} &=& -i[H,\rho ]-\kappa(1+\bar n_{th}) (a^{\dagger}a\rho -2a\rho a^{\dagger}+
\rho a^{\dagger}a)
\nonumber \\
 & & -\kappa \bar n_{th} (aa^{\dagger}\rho -2a^{\dagger}\rho a +
\rho aa^{\dagger})
\end{eqnarray}
describing the situation whenever a atom is present in the cavity [5]. $H$ is the 
Jaynes-Cummings Hamiltonian [2] in interaction picture given by
\begin{equation}
H=g(S^+a + S^-a^\dagger)
\end{equation}
with $g$ representing the strength of the atom-field couping. $a$
is the photon annihilation operator and $S^+$ and $S^-$ are the Pauli
pseudo-spin operators for the two-level system. As mentioned earlier, a atom takes 
a time $\tau$ to pass through the cavity. These atomic events are seperated 
by random durations, $t_{cav}$, during which the cavity evolves under its own 
dynamics.  Hence we set $H=0$ in Eq. (2) during $t_{cav}$. Processes like 
these atomic events seperated by random intervals are known as Poisson processes 
in literature encountered in various branches of physics, for example, 
radioactive materials emitting alpha particles. A sequence of durations of such 
processes can be obtained from uniform deviates, also called random numbers, $x$ 
generated using a computer such that $0~<~x~<1$, and then by using the 
relation [6]
\begin{equation}
t_R=-\mu\ln(x)
\end{equation}
where $t_R=t_{cav}+\tau$. $\mu~=~1/R$ where $R$ is the flux rate of atoms.  \\
\indent
We have carried out numerical simulation of the dynamics with the data taken 
from the experimental arrangements [1] in which $g=39~kHz$ and the $\tau =40~\mu s$ 
was one of the atom-field interaction times. This gives $g\tau = 1.56$, 
a condition required for generating a Fock state of $n_0$ photons where 
$n_0$ satisfies $\sin g\tau\sqrt{n_0+1}=0$ in an ideal cavity $(Q=\infty)$ [7]. 
Since the experimantal arrangements are close to ideal situation, it was hoped 
that such Fock states could be attained experimentally. Indeed, such results have 
been reported in Ref. 1. However, our numerical simulations [8] does not confirm 
these conclusions. Instead, it gives photon fields with very narrow distribution 
functions (sub-Poissoninan) centred about n. Figs. 1 and 2 display distribution 
function $P(n)$ narrowly centred about $n=14$.\\ 

\begin{figure}
\begin{center}
\includegraphics[width=6.0cm]{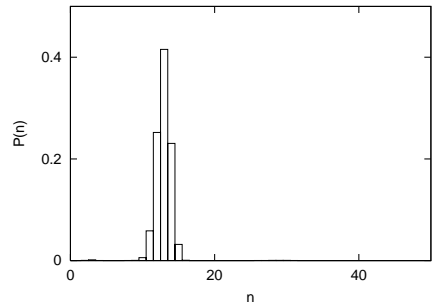}
\end{center}
{\caption{
Cavity photon distribution function at the exit of the 7000th atom.
}}
\end{figure}

\begin{figure}
\begin{center}
\includegraphics[width=6.0cm]{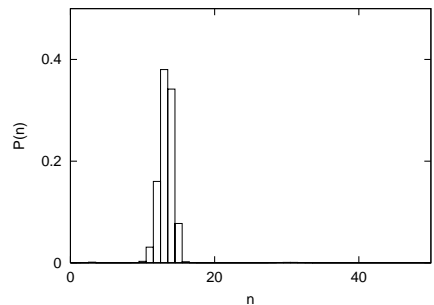}
\end{center}
{\caption{
P(n) vs n at the moment of the 9000th atom leaving the cavity.
}}
\end{figure}





The reason for these results is simple. The cavity dissipation, although very small, 
effects the coherent atom-field interaction and moreover the randomness in $t_{cav}$ 
makes the photon distribution function fluctuate all the time centred about $n_0$ in 
addition to making it broader.\\ 
\indent
In this experiment [1], the atoms coming out of the cavity are subjected to 
measuments from which state of the cavity field is inferred. The atoms enter 
the cavity in the upper $\vert a\rangle$ of the two states $\vert a\rangle$ and 
$\vert b\rangle$. The exiting atom is, in general, in a state 
\begin{equation}
\vert\psi\rangle = a\vert a\rangle + b\vert b \rangle
\end{equation}
with $p_a=\vert a\vert ^2$ and $p_b=\vert b\vert ^2$ are the probabilities of the 
atom being in the states $\vert a\rangle$ and $\vert b\rangle$ respectively.
According to the Copenhagen interpretation of quantum mechanics [9], this 
wave function {\it collapses} (or is projected) to either $\vert a\rangle$ or 
$\vert b\rangle$ the moment a measurement is made on it. Due to this inherent 
nature of quantum mechanics, a noise is associated with the measurement 
which is know as {\it quantum projection noise} [10]. We define the projection 
operator $J=\vert a\rangle\langle a\vert$. The variance in its measurement 
is given by
\begin{equation}
(\Delta J)^2=\langle J^2\rangle - \langle J\rangle^2=p_a(1-p_a)
\end{equation}
We find that $(\Delta J)^2=0$ only when $p_a=1~or~0$. For the generation of a 
Fock state, it is necessary that the atom should leave the cavity unchanged in its upper 
state [3,4,7]. Hence, for such a situation we must have $p_a=1$ in which case 
$(\Delta J)^2$ should be 0. We find from our numerical simulations 
that that $p_a\equiv P(a)$ is mostly about 0.8 [Fig. 3] and, hence, 
$(\Delta J)^2\ne 0$ always. This obviously indicates that the cavity 
field is in a linear superposition of Fock states giving a photon distribution 
function with the normalized variance $v~>~0$ (For a Fock state $v=0$). 
Indeed, we find that the $v$ is about $0.5$ in our calculations, presented in Fig. 4, 
indicating a sub-Poissonian nature of the cavity field. By itself, it carries a 
signature of quantum mechanics. We further notice in Fig. (4) that there are small 
fluctuations in $v$ due to the fluctuations in $P(n)$ [Figs 1 and 2]. Also, $v$ is 
nowhere near $0$ in Fig. 4.\\

\begin{figure}
\begin{center}
\includegraphics[width=12.0cm]{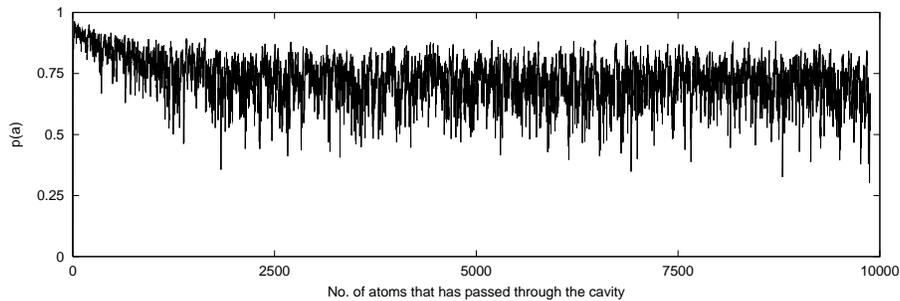}
\end{center}
{\caption{Population of the upper state of the individual atoms at the exit from
the cavity.}}
\end{figure}




\begin{figure}
\begin{center}
\includegraphics[width=6.0cm]{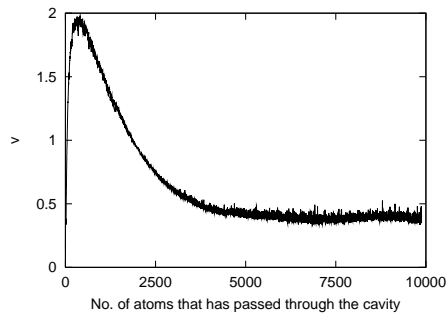}
\end{center}
{\caption{
Fluctuations in $v$ at the exit of successive atoms from the cavity.
}}
\end{figure}
 

The field ionization techniques used in Ref. [1] to detect the outgoing 
atomic states , obviously, cannot incorporate the above quantum projection 
noise. The photon statistics inferred from the 
measured atomic statistics would then be correct only to the extent one could 
afford to neglect the quantum noise. But, our observations in Figs. 1-4 clearly 
show that this is crucial for the generation of a photon Fock state. In other 
words, the situation $\Delta J=0$ just does not happen due to the non-stop 
dissipation of the cavity field and also due to the randomness in $t_{cav}$.
Further, as mentioned earlier, the small but finite $\bar n_{th}$ in the 
equation of motion (Eq. 2) has a major influence on this dissipation [5].
 Hence, the analysis of the micromaser dynamics in Ref. 11 does not show 
cavity field dissipation clearly since the effects of the finite cavity 
temperature has not been properly included there. \\
\indent
We have carried out simulation until about ten thousand atoms passed through 
the cavity and carrying out the simulations any further would only be a 
repitition of the fluctuations in Figs. 1-4. In other words, the nagging reservoir 
dynamics does not allow the cavity field to settle down to a photon number state. 
A similar conclusion can also be inferred in a recent report [12] where the authors 
show that Fock states are fragile in thermal baths.  \\ 
\noindent
{\bf Acknowledgement:}\\
I am indebted to Professor R. K. Bullough of Manchester, England for his 
neverending support without which the study of micromaser would not have 
reached this conclusion.\\ \\
{\bf References:}\\
\noindent
[1] B. T. H. Varcoe, S. Brattke, M. Weldinger, and H. Walther, 
Nature, {\bf 403}, 743 (2000);S. Brattke, B. T. H. Varcoe, and H. Walther, 
Phys. Rev. Lett. {\bf 86}, 3534 (2001). \\
\noindent
[2] E. T. Jaynes and F. W. Cummings, Proc.IEEE {\bf 51}, 89 (1963).\\
\noindent
[3] N. Nayak, Opt.Commun. {\bf 118}, 114 (1995).\\
\noindent
[4] P. Filipowicz, J. Javanainen, and P. Meystre, Phys. Rev. A{\bf 34}, 
3077 (1986).\\
\noindent
[5] N. Nayak, J. Opt. Soc. Am. B {\bf 13}, 2099 (1996).\\
\noindent
[6] D. E. Kunth, {\it The Art of Computer Programming} (addison-Wesley, Reading, 
Mass. 1981), Vol.2.\\
\noindent
[7] A. K. Rajagopal and F. W. Cummings, Phys.Rev. A {\bf 39}, 3414 (1989).\\
\noindent
[8] All the numerical results have been evaluated with an error of 1 part in 
$10^{12}$.\\
\noindent
[9] Max Jammer, {\it The Conceptual Developement of Quantum Mechanics} (McGraw-Hill, 
1966), Ch.7. \\
\noindent
[10] W. M. Itano, J. C. Bergquist, J. J. Bollinger, J. M. Gilligan, 
D. J. Heinzen, F. L. Moore, M. G. Raizen, and D. J. Wineland, 
Phys. Rev.  A {\bf 47}, 3554 (1993).\\
\noindent
[11] S. Brattke, B. Englert, B. T. H. Varcoe, and H. Walther, 
J. Mod. Opt. {\bf 47}, 2857 (2000).\\
\noindent 
[12]A. Serafini, S. De Siena, and F. Illuminati, Mod. Phys. Lett. {\bf 18}, 687 
(2004). 
\end{document}